\documentclass[aps,prd,superscriptaddress,nofootinbib,onecolumn,12pt]{revtex4-2}

\usepackage{amsmath,amssymb,amsbsy,amstext,amsfonts,mathrsfs,latexsym,bm,physics}
\usepackage[colorlinks=true,urlcolor=blue,anchorcolor=blue,citecolor=blue,linkcolor=blue,linktocpage=true,pdfa=true]{hyperref}
\usepackage{graphicx}

\begin{document}

\title{NUTs, Bolts and Stokes Phenomena in the \\ No-Boundary Wave Function}

\author{Jean-Luc Lehners}
%\email{jean-luc.lehners@aei.mpg.de}
\affiliation{Max Planck Institute for Gravitational Physics (Albert Einstein Institute) \\ D-14476 Potsdam, Germany}

\begin{abstract}
%\vspace{0.3cm}
In this note, we revisit and extend the analysis of the no-boundary wave function for the minisuperspace model in which the universe is described by a biaxial Bianchi~IX metric. As matter content, we simply assume a positive cosmological constant. We find that two Stokes phenomena occur, at large squashing parameters of the spatial section of the universe. These Stokes phenomena eliminate potentially dominant Taub-Bolt-de Sitter saddle point geometries and are crucial for the consistency of the model. They also imply that phase transitions occur at certain levels of squashing, where NUT and Bolt saddle points exchange dominance. 
\end{abstract}

\maketitle

%\newpage 

\tableofcontents

\section{Introduction}

Perhaps the most intriguing approach to quantum theory is Feynman's path integral in which, in order to calculate a transition amplitude for a particle to propagate from A to B, we are instructed to sum over all possible paths connecting the starting and end points. Once gravity is included, the role of the particle paths is played by entire spacetimes and matter configurations living on them, interpolating between initial and final field configurations. Defining such path integrals precisely and rigorously remains a largely unfinished task, however. This is not only due to the non-renormalisability of general relativity, but also to many conceptual questions, such as: which spacetime manifolds should be allowed in the sum, and which not? How regular should the manifolds be? Should one sum over topologies, or fix the topology? Which boundary conditions should one impose on the fields? Many works keep exploring these topics, see {\it e.g.} 
\cite{Honda:2024aro,Ailiga:2023wzl,Dittrich:2023rcr,Matsui:2023tkw,Gielen:2022yez,Matsui:2022lfj,Jia:2022nda,Isichei:2022uzl,Lehners:2022mbd} for recent examples.

Some of these questions are not only mathematical in nature, but potentially reflect physical realities. A particularly influential proposal, addressing some of these questions, is the no-boundary proposal of Hartle and Hawking \cite{Hawking:1981gb,Hartle:1983ai} (for a review see \cite{Lehners:2023yrj}). These authors suggest that in order to calculate amplitudes for present day events, one should sum over manifolds that do not contain a boundary to the past. The lack of a past boundary is meant to eliminate the need for specifying unknowable conditions in our far past, thus providing a prescription for calculating the ``wave function of the universe''. Included in this prescription is the idea that one should sum over no-boundary manifolds with various topologies. 

Attempts to define the no-boundary proposal more precisely in concrete (typically symmetry-reduced) models have a long history, for a few examples see {\it e.g.} \cite{Halliwell:1988ik,Halliwell:1990tu,Daughton:1998aa,Hartle:2008ng,Feldbrugge:2017kzv}. Here we wish to illustrate these issues with a simple, yet not too simple, cosmological model, in which the universe is described by a biaxial Bianchi IX metric, and the matter content is a positive cosmological constant $\Lambda.$ This type of metric has spatial sections that are squashed $3-$spheres, with a single parameter controlling the amount of squashing. The squashing can be pictured as the relative ratio of the sizes of a circle and a $2-$sphere, making up the $3-$sphere as a fibration. The model is useful in that it can be treated analytically to a large extent, yet it is non-trivial in that it reveals many of the issues mentioned above. This kind of model has been studied before, most recently in \cite{DiazDorronsoro:2018wro,Feldbrugge:2018gin,Janssen:2019sex}. In \cite{DiazDorronsoro:2018wro} the contributions of Taub-NUT-de Sitter geometries, which furnish a subset of Bianchi IX metrics, were analysed. There it was shown that they are stable (with damped perturbations around them) provided one imposes an appropriate condition on the initial expansion rate of the $2-$sphere direction and provided one chooses an appropriate sum over complex metrics. In \cite{Feldbrugge:2018gin}, it was shown that imposing instead that the universe starts at zero scale factor, and summing over Lorentzian metrics, does not lead to physically sensible results, despite the apparently physically sensible assumptions. One is then left with the choice of either abandoning the no-boundary proposal, or accepting a sum over complex metrics as a fundamental feature. We will adopt the latter approach here. And in \cite{Janssen:2019sex}, the results of \cite{DiazDorronsoro:2018wro} were extended to include contributions from Taub-Bolt-de Sitter geometries, which admit a different topology. There it was shown again that certain conditions should be imposed on the initial expansion rates, and that the path integral should be defined as a sum over a subset of complex metrics. This last work showed that it seemed plausible that different topologies should dominate at different values of the squashing.

In the present work, we continue the analysis started in \cite{Janssen:2019sex} by showing exactly how different topologies come to dominate in different regimes. In fact, these phase transitions can be traced back to Stokes phenomena, which cause various saddle points of the gravitational path integral to become irrelevant beyond certain (squashing) parameter ranges. Stokes phenomena, discovered in 1847 by G.G. Stokes, reveal that the asymptotic expansion of an analytic function can vary suddenly as a parameter is changed. In the present context, this has the consequence of suppressing certain Bolt geometries, letting NUT geometries become dominant in various parameter ranges. 

Our work supports the idea that a sum over topologies is not only natural, but also leads to concrete physical consequences. Moreover, the Stokes phenomena help in making the wave function normalisable, at least up to the level of precision that can currently be calculated reliably. That said, allowing for different topologies in quantum gravity is also known to lead to puzzles, especially those associated with the effects of wormholes \cite{Hebecker:2018ofv,Jonas:2023ipa,Jonas:2023qle}. Although this work has no direct bearing on these issues, it reinforces the paradoxes just mentioned and motivates a closer investigation thereof.

The plan of the paper is as follows: in section \ref{sec:PL} we will briefly review the Picard-Lefschetz method of treating Feynman-type path integrals, and review how Stokes phenomena arise in this approach. Then, in section \ref{sec:lapse} we introduce the biaxial Bianchi IX model and solve it analytically to the extent that this is feasible. In section \ref{sec:sad}, we use numerical methods to determine which kinds of saddle points contribute to the gravitational path integral, and exhibit the general properties of the resulting no-boundary wave function. We also briefly investigate the consequences for the closely related tunneling proposal in section \ref{sec:tunnel}. Finally, in section \ref{sec:disc}, we conclude.

\section{Picard-Lefschetz theory and Stokes phenomena} \label{sec:PL}

In quantum cosmology one regularly encounters integrals which are of Feynman type, 
\begin{align}
\label{eq:integral}
\Psi = \int_{\cal C} \mathrm{d}x\, e^{\frac{i}{\hbar}S[x]}\,,
\end{align}
where the integrand is an oscillating function. Here, at least at first, we will assume that  (the action) $S[x]$ is a real function of the variable $x$, and $\hbar$ is a real parameter. In applications below $x$ will be the lapse function $N.$ The domain of integration ${\cal C}$ remains unspecified at this point. Due to the oscillations in the integrand, the analysis of such integrals requires some care. The proper mathematical framework for dealing with such integrals is called Picard-Lefschetz theory -- a thorough review can be found in \cite{Witten:2010cx}, applications to quantum mechanics in \cite{Tanizaki:2014xba} and to gravity in \cite{Feldbrugge:2017kzv}. Here we will restrict ourselves to a few features salient for the present paper.

The main idea is to let the variable $x$ become complex, and to deform the integration contour in such a way as to eliminate the oscillations of the integrand, thereby rewriting the oscillating integral as a sum over manifestly convergent integrals. Let us see in more detail how this can be achieved. First expand the meaning of $S[x]$ by considering it to be a holomorphic function of the now complex-valued $x.$ It is useful to write $x=u^1+i u^2$ for real $u^1, u^2,$ and to similarly decompose the exponent as $\mathcal{I}=iS/\hbar = W + i P.$ We will call $W$ the \emph{weighting} and $P$ the \emph{phase}. Note that when $S$ is real, the weighting is constant (and zero) and only the phase is changing. Convergent integrals, such as a standard Gaussian, have the opposite behaviour: they have a constant phase and an asymptotically decreasing weighting. In this spirit, let us define a complex curve of downward flow via
\begin{equation}
\frac{\mathrm{d}u^i}{\mathrm{d}\lambda} = -g^{ij}\frac{\partial W}{\partial u^j}\,,
\label{eq:dw}
\end{equation}
where $\lambda$ is a parameter along the flow and $g_{ij}$ is a Riemannian metric on the complex plane, which we will simply take to be the trivial metric\footnote{The trivial metric is $\mathrm{d}s^2 = |\mathrm{d}x|^2.$ With complex coordinates, $(u,\bar{u})=\bigl(u^1+i u^2,u^1-i u^2\bigr)$, we have $g_{uu}=g_{\bar{u}\bar{u}}=0,\,g_{u\bar{u}}=g_{\bar{u}u}=1/2$.}. By calculating  $\frac{\mathrm{d}W}{\mathrm{d} \lambda} = \sum_i\frac{\partial W}{\partial u^i}\frac{\mathrm{d}u^i}{\mathrm{d}\lambda} = -\sum_i\left(\frac{\partial W}{\partial u^i}\right)^2<0,$ one can see that indeed the weighting decreases along this flow. 

Since $W=(\mathcal{I}+\bar{\mathcal{I}})/2,$ we can also write the downwards flow equations as
\begin{equation}
\frac{\mathrm{d}u}{\mathrm{d}\lambda} = - \frac{\partial {\bar{\cal I}}}{\partial \bar{u}}, \quad \frac{\mathrm{d}\bar{u}}{\mathrm{d}\lambda} = - \frac{\partial {{\cal I}}}{\partial {u}}\,.
\end{equation} 
By analogy, one can define upwards flow as
\begin{equation}
\frac{\mathrm{d}u}{\mathrm{d}\lambda} = + \frac{\partial {\bar{\cal I}}}{\partial \bar{u}}, \quad \frac{\mathrm{d}\bar{u}}{\mathrm{d}\lambda} = + \frac{\partial {{\cal I}}}{\partial {u}}\,.
\end{equation} 
Note that the upwards/downwards flows end at either a singularity where the weighting diverges to plus/minus infinity, or at saddle points of the integrand where the action is stationary. We may thus conclude that (equal numbers of) steepest ascent and descent lines radiate out from saddle points, until they reach infinite (negative or positive) weighting. Steepest descent contours are also known as ``Lefschetz thimbles'' and will be denoted $\cal{J}_\sigma,$ while steepest ascent contours are denoted $\cal{K}_\sigma.$ When there is no degeneracy, they intersect only at the saddle point $\sigma$ which they are associated with,
\begin{equation}
{\rm Int}({\cal J}_\sigma, {\cal K}_{\sigma'})=\delta_{\sigma \sigma'}\,.\label{eq:intersection}
\end{equation}

The steepest ascent/descent contours share a useful property, namely that the phase is constant along them:
\begin{equation}
\label{eq:imh}
\frac{\mathrm{d} P}{\mathrm{d}\lambda} = \frac{1}{2i}\frac{\mathrm{d}({\cal I} - \bar{\cal I})}{\mathrm{d}\lambda} = \frac{1}{2i}\left( \frac{\partial {\cal I}}{\partial u}\frac{\mathrm{d}u}{\mathrm{d}\lambda} - \frac{\partial \bar{\cal I}}{\partial \bar{u}}\frac{\mathrm{d}\bar{u}}{\mathrm{d}\lambda}\right) = 0\,.
\end{equation}
This property helps one to easily find such contours in practice, by starting from saddle points and following the loci of constant phase. Hence we see that ``steepest descent'' and ``stationary phase'' methods are in fact one and the same thing.

We are now ready to rewrite the integration contour as a sum over thimbles, in mathematical language
\begin{equation}
\label{eq:contourexp}
{\cal C} = \sum_\sigma n_\sigma {\cal J}_\sigma\,,
\end{equation}
where $n_\sigma \in (0,\pm 1)$ and where the sign depends on the (arbitrarily chosen) orientation of the contours. Taking the intersection with $\cal{K}_\sigma$ on both sides and using Eq.~\eqref{eq:intersection}, we see that 
\begin{align}
n_\sigma= {\rm Int}(\mathcal{C}, {\cal K}_{\sigma})\,.
\end{align}
This equation tells us that the integration contour $\cal{C}$ can be rewritten as a sum over thimbles associated with saddle points that have a lower weighting than that of the integrand evaluated along $\cal{C},$ since one needs to move along a steepest ascent line from the saddle to the integration contour in order for the saddle point to be relevant. Now we are in a position to rewrite the original integral in a most useful form
\begin{align}
\Psi = \int_{\cal C} \mathrm{d}x \, e^{i S[x]/\hbar} & = \sum_\sigma n_\sigma \int_{{\cal J}_\sigma} \mathrm{d} x \, e^{\frac{i}{\hbar}S[x]} \\
 & = \sum_\sigma n_\sigma \, e^{i \, P(x_\sigma)}\int_{{\cal J}_\sigma} e^W \mathrm{d}x \\ 
& \approx \sum_\sigma n_\sigma \, e^{i S(x_\sigma)/\hbar}\left[A_\sigma+\mathcal{O}(\hbar)\right]\,.
\end{align} 
The first line above is an exact rewriting. Since the phase is constant along thimbles, it can be pulled out, as done in the second line, which is still exact. Then, in the third line, we may use the fact that in applications $\hbar$ is small, so that we may approximate the integral by a sum of values at the relevant saddle points, with $A_\sigma$ being prefactors that can be obtained by integrating over fluctuations around the saddle points.

\begin{figure}[ht]
	\centering
	\includegraphics[width=0.95\textwidth]{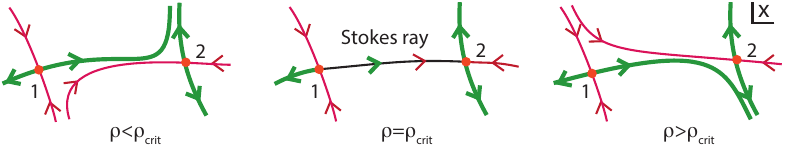}
	\caption{As a parameter of an integral is varied, two saddle points may obtain the same phase at a critical value $\rho_{crit}$ and become linked via a Stokes ray. Passing such a critical value, the flow lines undergo a topological change (compare the left and right graphs), known as a Stokes phenomenon. The figure illustrates this by showing saddle points in orange, steepest descent lines in green and ascent lines in red, in the complex $x$ plane. The Stokes ray, which is a steepest descent contour from the point of view of saddle $1$, and a steepest ascent contour as seen by saddle $2,$ is drawn in black. The arrows indicate the direction of decreasing weighting.}
	\label{fig:stokes}
\end{figure}

So far, we assumed that there was no degeneracy. There are two sources of degeneracy that we need to pay attention to, however. These can arise if the integrand has certain symmetry properties, or when a parameter in the integral is varied and passes through a critical value. The first type of degeneracy is when saddle points happen to coalesce. This gives rise to a saddle point of higher order, where not only the first derivative of the action vanishes, but also the second (or higher) derivative(s). Evidently, at such a degeneracy, the saddle points obtain both the same weighting and the same phase.

A second type of degeneracy is when two saddle points obtain the same phase, though they are not coincident and don't share the same weighting. In such a case, it can happen that the steepest descent line from one saddle reaches the second saddle. From the second saddle's point of view, this is then a steepest ascent line. This connecting flow is known as a Stokes ray. When the parameter in question is varied further, a topological change in the flow lines occurs (see Fig.~\ref{fig:stokes} for an illustration). This is known as a Stokes phenomenon and, as we will see, it can affect the relevance of the saddles to the integral. The concrete examples below will illustrate this phenomenon, with the parameter in question being the anisotropy of the universe.

\section{Biaxial Bianchi IX model and lapse integral} \label{sec:lapse}

Before looking at the anisotropic case, let us briefly review how one constructs the no-boundary wave function when the assumed spatial geometry of the universe is homogeneous and isotropic. We will focus on the issues that will be relevant to the anisotropic calculation performed below -- for more details of the isotropic calculation see \cite{DiTucci:2019dji,DiTucci:2019bui}. We will be interested in a theory of gravity coupled to a cosmological constant $\Lambda,$ with action 
\begin{eqnarray}
S = \int_{\cal M} \mathrm{d} ^4 x  \sqrt{-g} \left( \frac{R}{2} - \Lambda \right)  + \int_{\partial {\cal M}_{1}} \mathrm{d}^3 y \sqrt{h}K \,. \label{LorentzianActionND}
\end{eqnarray}
Although one expects higher order terms to be present as quantum gravitational corrections, these terms play a sub-leading role as long as the curvature remains significantly below the Planck or string scale, as will be the case here \cite{Jonas:2020pos}. Also, other matter types can be included, again without significantly affecting our results \cite{Lehners:2023yrj}. What is important is that we are including the Gibbons-Hawking-York (GHY) boundary term, with induced metric $h$ and trace of the extrinsic curvature $K,$ but only on the final hypersurface \cite{York:1972sj,Gibbons:1976ue}. This is because we would like the variational problem to be well-defined when imposing a Dirichlet boundary condition (fixing the scale factor of the universe) there. Since our initial condition is the no-boundary initial condition, we will not impose an initial boundary term, which would go against the spirit of the no-boundary idea \cite{Louko:1988bk}. As we will see momentarily, this naturally leads to a Neumann boundary condition for the path integral.

In order to implement the no-boundary idea, it is necessary to consider the universe to be spatially closed. We will write the closed Robertson-Walker metric in the useful form
\begin{align} \label{metriciso}
\mathrm{d}s^2 = - \frac{N^2}{q}\mathrm{d}t^2 + \frac{q}{4} (\sigma_1^2+\sigma_2^2+ \sigma_3^2)\,,
\end{align}
where $N$ is the lapse function, $q(t)$ the scale factor and the one-forms on the sphere are given by $\sigma_1 = \sin\psi \mathrm{d}\theta - \cos \psi \sin \theta \mathrm{d}\varphi$, $\sigma_2 = \cos \psi \mathrm{d}\theta + \sin \psi \sin \theta \mathrm{d} \varphi$, and $\sigma_3 = \mathrm{d}\psi + \cos\theta \mathrm{d}\varphi$ with coordinate ranges $0 \leq \psi \leq 4 \pi$, $0 \leq \theta \leq \pi$, and $0 \leq \varphi \leq 2 \pi.$ The time coordinate can be chosen to span the interval $[0,1].$ With this choice of metric, the action simplifies considerably and becomes quadratic in $q$ \cite{Halliwell:1988ik}
\begin{align}
     S_q=2\pi^2\int_0^1 \mathrm{d}t \left[-\frac{3}{4N}\dot{q}^2 + N(3-\Lambda q)\right] - \frac{3\pi^2}{N}q\dot{q}|_{t=0}\,, \label{eq:act}
\end{align}
where dots indicate time derivatives. Note that integrations by parts have eliminated the surface term at $t=1$ and introduced one at $t=0.$ Variation of the action with respect to $q$ then allows one to consistently fix the scale factor on the final hypersurface, $q(t=1)=q_1,$ while allowing one to fix the momentum $\dot{q}/N$ at $t=0.$ But what should it be fixed to? The Friedmann equation following from \eqref{eq:act} is 
\begin{align}
    \frac{\dot{q}^2}{4N^2} + 1  = \frac{\Lambda}{3} q\,.
\end{align}
If we would like the universe to close off in a regular manner, then when $q=0$ the equation above implies that we should set $\dot{q} = \pm 2Ni.$ The choice of sign is rather crucial here: as any no-boundary geometry has a Euclidean cap, one can think of the sign above as a choice of Wick rotation. The positive sign ensures that fluctuations are stable (and obtain the canonical sign as gravity is turned off), while a negative sign leads to divergences. We will thus fix
\begin{align}
    \frac{\dot{q}}{2N}(t=0) = +i\,. \label{regularityq}
\end{align}

With these mixed Neumann-Dirichlet boundary conditions, the no-boundary wave function is a function of the final scale factor $q_1$ only, and can be expressed as the path integral
\begin{align}
    \Psi_{ND}(q_1) = \int_{\dot{q}=2Ni}^{q=q_1} {\cal D}q \int_{\cal C} \mathrm{d}N \, e^{\frac{i}{\hbar}S_q}\,.
\end{align}
With the appropriate choice of integration contour, there are two saddle points $N_\sigma$ that contribute to the integral (these differ only by the sign of $\textrm{Re}(N_\sigma)$), leading to the (real) wave function
\begin{align}
    \Psi_{ND}(q_1) \approx e^{+\frac{12\pi^2}{\hbar \Lambda} - i \frac{12\pi^2}{\hbar\Lambda}\left(\frac{\Lambda}{3} q_1 - 1\right)^{3/2}} + \, e^{+\frac{12\pi^2}{\hbar \Lambda} + i \frac{12\pi^2}{\hbar\Lambda}\left(\frac{\Lambda}{3} q_1 - 1\right)^{3/2}}\,, \label{wfND}
\end{align}
Note that the weighting $\frac{12\pi^2}{\hbar \Lambda}$ is constant (and favours small $\Lambda$) while the phase grows asymptotically in proportion to the volume $q_1^{3/2}$ of the universe. Each contribution is thus of Wentzel-Kramers-Brillouin (WKB) form, explaining the origin of a classical spacetime \cite{Hartle:2008ng}.

We can now expand these results to an anisotropic minisuperspace model previously studied in \cite{DiazDorronsoro:2018wro,Feldbrugge:2018gin,Janssen:2019sex} and known as biaxial Bianchi IX (or dubbed BB9 in \cite{Janssen:2019sex}, presumably as a nod to the Star Wars droid BB8). If one thinks of a $3-$sphere as a fibration of a circle $S^1$ over a sphere $S^2,$ then the  BB9 metric can be pictured as giving different radii to the $S^1$ and $S^2$, in other words it describes a squashed $3-$sphere (see \cite{Daughton:1998aa}), 
\begin{align} \label{metricbb9}
\mathrm{d}s^2 = - \frac{N^2}{q}\mathrm{d}t^2 + \frac{p}{4} (\sigma_1^2+\sigma_2^2) + \frac{q}{4} \sigma_3^2\,,
\end{align}
where $q(t),p(t)$ are the time dependent radii of the circle and sphere, respectively. Sometimes, this is also called the Taub spacetime \cite{Ryan:1975jw}. We will be interested in two different versions, with differing topologies: the Taub-NUT space in which both $q$ and $p$ reach zero simultaneously, and the Taub-Bolt space in which the circle radius $q$ reaches zero while the $2-$sphere radius $p$ remains non-zero. Both versions have zero spatial volume $2\pi^2 q^{1/2}p$ at the origin, and both are regular there, thus providing suitable saddle points for the no-boundary wave function.

When evaluated on this metric, the action \eqref{LorentzianActionND} reduces to
\begin{align}
    S=&2\pi^2 \int_0^1 \dd t \left[N\left(4-\frac{q}{p} - p\Lambda \right) +\frac{1}{4N}\left(-\frac{q}{p}\dot{p}^2 + 4 q \ddot{p} + 2 p \ddot{q} + 4 \dot{p} \dot{q}\right)\right] \nonumber \\ & + \frac{\pi^2}{N}\left(-p\dot{q}-2q\dot{p} \right)\mid_{t=1}\,,
\end{align}
where the second line consists of the boundary contribution on the final hypersurface at $t=1.$ Varying the action with respect to the fields yields the equations of motion
\begin{align}
    2p\ddot{p} -\dot{p}^2 & = 4N^2\,, \label{eomp} \\
    \ddot{q} + \frac{\dot{p}}{p}\dot{q} & = N^2 \left(2\Lambda - 4\frac{q}{p} \right)\,,
\end{align}
supplemented by the boundary variations
\begin{align}
    \left( \Pi_q \delta q + \Pi_p \delta p\right)\mid_{t=1} + \left( q \delta \Pi_q + p \delta \Pi_p\right)\mid_{t=0}\,, \label{bdyvariation}
\end{align}
which we have written in terms of the canonical momenta
\begin{align}
    \frac{1}{2\pi^2}\Pi_q = -\frac{1}{2N}\dot{p}\,, \qquad \frac{1}{2\pi^2}\Pi_p = -\frac{1}{2N}\left( \dot{q} + \frac{q}{p}\dot{p}\right)\,.
\end{align}
Before discussing boundary conditions let us point out a peculiarity of this model, which is that the action is linear in $q$ and thus the path integral over $q$ results in a (complex) functional delta function, imposing Eq.~\eqref{eomp}. In other words, the path integral only has support on solutions of the equation of motion for $p.$ With this substitution we obtain
\begin{align}
    S=2\pi^2 \int_0^1 \dd t N\left(4- p\Lambda \right) \,\, + \frac{\pi^2}{N}\left[-q\dot{p}\mid_{t=1} - (q\dot{p}+ p \dot{q})\mid_{t=0} \right]\,.
\end{align}
Note that the action depends on $q$ only via its boundary values. 

To proceed, we must impose appropriate boundary conditions. On the final hypersurface we will simply impose Dirichlet conditions $q(t=1)=q_1$ and $p(t=1)=p_1.$ On the initial hypersurface, we would like to implement the no-boundary idea, meaning that we would like the saddle point contributions to be compact and regular. However, in quantum theory we cannot simultaneously impose a condition on a field and its conjugate momentum. Eq.~\eqref{bdyvariation} then allows for $4$ possible choices {\it a priori}: 1. Fixing $q_0=p_0=0.$ 2. Fixing $p_0=0$ and $\Pi_{q,0}=-2\pi^2i$ ({\it cf.} Eq. \eqref{regularityq}) 3. Fixing $q_0=0$ and $\Pi_{p,0}=-2\pi^2i.$ 4. Fixing both momenta $\Pi_{q,0}=\Pi_{p,0}=-2\pi^2 i.$ Choice $1$ is plagued by a sign ambiguity related to the effective Wick rotation imposed. This choice would thus include metrics with unstable (ghost-like) excitations \cite{Feldbrugge:2017fcc,Feldbrugge:2017mbc}. Moreover, the condition $p_0=0$ would eliminate Taub-Bolt metrics \cite{Janssen:2019sex}. For choices $2$ and $3,$ the sign of the Wick rotation is fixed by the sign of the momentum \cite{DiTucci:2019dji}. Meanwhile, choice $4$ is overconstraining \cite{Halliwell:1990tu}. Thus the two interesting choices are $2$ and $3,$ and it seems natural that one should sum over both. In other words, the no-boundary path integral should presumably include a sum over boundary conditions, allowing a sum over topologies. Our results below will support this point of view.

For completeness, we will write out the general solutions to the equations of motion with $p(0)=p_0, \, p(1)=p_1, \, q(0)=q_0, \, q(1)=q_1$ \cite{Anabalon:2018rzq,Janssen:2019sex}:
\begin{align}
    p(t)& =p_0 + 2\left(\sqrt{p_0p_1 - N^2}-p_0\right)t+  \left(p_0+p_1-2 \sqrt{p_0p_1-N^2}\right)t^2 \label{solp}\\
    q(t) & = \frac{\Lambda}{p(t)}  \left(\frac{p_0q_0}{\Lambda }+ t \frac{\left(p_0 \left(-2 \sqrt{p_0 p_1-N^2} \left(N^2+\frac{3 q_0}{\Lambda }\right)-N^2 p_1+\frac{3 p_1 (q_0+q_1)}{\Lambda }\right)\right)}{3 \sqrt{p_0 p_1-N^2}}  \right. \nonumber \\ & \left. +t^2\frac{2 N^4  -p_0 p_1 \left(N^2+\frac{3 (q_0+q_1)}{\Lambda }\right)}{3 \sqrt{p_0 p_1-N^2}} +p_0 \left(\frac{5}{3} N^2+\frac{q_0}{\Lambda }\right)+p_1  \left(\frac{ q_1}{\Lambda }-\frac{1}{3}N^2\right)\right. \nonumber \\ & \left. +\frac{4}{3} t^3  N^2 \left(\sqrt{p_0 p_1-N^2}-p_0\right) \right. \nonumber \\ & \left. +\frac{1}{3} t^4 N^2 \left(-2 \sqrt{p_0 p_1-N^2}+p_0+p_1\right)\right) \label{solq}
\end{align}
We will fix a branch by taking $\sqrt{p_0p_1-N^2}\to +iN$ as $p_0\to 0.$ The solutions above will allow us to determine the saddle point geometries. Their shape will depend on the boundary conditions imposed. Hence, in the following, we will treat both boundary conditions (labelled $2$ and $3$ above) in turn.

\section{Anisotropic no-boundary wave function} \label{sec:sad}

\subsection{Taub-NUT saddles}

\begin{figure}[ht]
	\centering
	\includegraphics[width=0.6\textwidth]{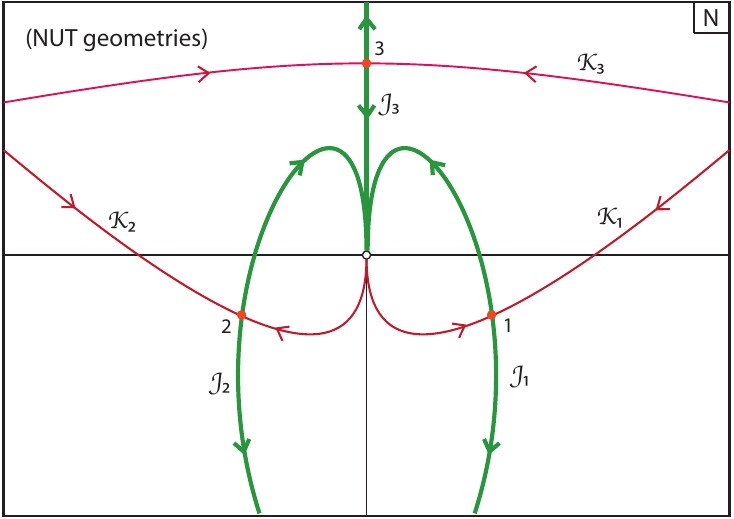}
	\caption{Saddle points and flow lines for NUT boundary conditions. The labelling is identical to that in Fig.~\ref{fig:stokes}.}
	\label{fig:posHHNUT}
\end{figure}

The first set of boundary conditions that we will impose are choice $2$ above, namely that $p_0=0$ and $\Pi_{q,0}=-2\pi^2i.$ Plugging these conditions into the solutions Eqs.~\eqref{solp} and \eqref{solq} and integrating over time, we obtain the action 
\begin{align}
    \frac{1}{2\pi^2}S_{NUT}(N)= -\frac{q_1p_1}{N} + i q_1 + (4-\frac{\Lambda}{3}p_1)N-i\frac{\Lambda}{3}N^2 \,,
\end{align}
which is now a function of the lapse $N$ only. This case has been analysed in detail in \cite{DiazDorronsoro:2018wro,Feldbrugge:2018gin,Janssen:2019sex}, for example, and proceeds rather similarly to the isotropic case reviewed above. The saddle points and flow lines are shown in Fig.~\ref{fig:posHHNUT} for final conditions in which the final scale factors are bigger than the cosmological scale $\sqrt{3/\Lambda}.$ 

\begin{figure}[ht]
	\centering
	\includegraphics[width=0.45\textwidth]{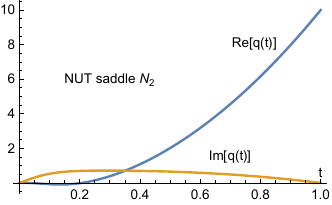}
 \includegraphics[width=0.45\textwidth]{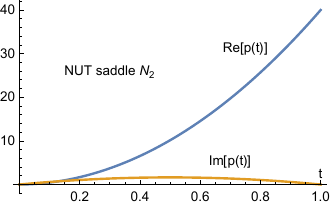}
	\caption{An example of a saddle point geometry for $\Lambda=3$ and final conditions $q_1=10, \, p_1=40.$ Note that the scale factors both start out at zero size and reach a real final value at $t=1.$}
	\label{fig:posHHsadNUT}
\end{figure}

There are three saddle points, one of which is purely Euclidean and is inappropriate for describing an evolving universe. The other two saddle points form a complex conjugate pair of geometries that are regular and that become Lorentzian asymptotically (at large scale factors). These are complex versions of the Taub-NUT-de Sitter spacetime. An example of a saddle point geometry is shown in Fig.~\ref{fig:posHHsadNUT}.

In order for these saddle points to be picked up by the path integral, the integration contour for the lapse must be chosen to include their thimbles, {\it i.e.} the integration contour must be ${\cal J}_1 \pm {\cal J}_2.$ Neither a Euclidean nor a Lorentzian integration contour are possible, but perhaps this is not so surprising given the complex boundary conditions that we had to impose. We will continue our discussion of the NUT saddles after including the second set of boundary conditions we are interested in.

\subsection{Taub-Bolt saddles}

We may also impose what we called choice $3$ for the boundary conditions above, namely imposing $q_0=0$ and $\Pi_{p,0}=-2\pi^2i,$ which in this case amounts to setting $\dot{q}(0)=2Ni.$ Obtaining the latter condition fixes the initial value of $p$ to 
\begin{align}
    p_0 = \frac{p_1 (\Lambda N^2-3q_1)^2}{4 N^2  (\Lambda  N+3 i)^2} + \frac{N^2 }{p_1}\,.
\end{align}
Plugging the boundary conditions into the equations of motion \eqref{solp}, \eqref{solq} and integrating over time, we obtain the lapse action 
\begin{align}
    \frac{1}{2\pi^2}S_{Bolt} = \left(4-\frac{\Lambda}{3} \right)N -\frac{p_1 q_1}{N}-\frac{4 \Lambda ^2 N^6+24 i \Lambda  N^5-N^4 \left(\Lambda ^2 p_1^2+36\right)+6 \Lambda  N^2 p_1^2 q_1-9 p_1^2 q_1^2}{12 N^2 p_1 (\Lambda  N+3 i)}\,,
\end{align}
which was presented in \cite{Janssen:2019sex} for $\Lambda=1.$ The last term is obtained after taking square roots, and care must be taken in fixing its overall sign. One may also write the action in the somewhat simpler form
\begin{align}
    \frac{1}{2\pi^2}S_{Bolt}(N)= \left(4-\frac{\Lambda}{3} \right)N -\frac{p_1 q_1}{N}-\frac{ N^2(\Lambda N + 3i)}{3 p_1} +\frac{ p_1 (\Lambda N^2 -3q_1)^2}{12 N^2  (\Lambda  N+3 i)}\,, \label{Boltaction}
\end{align}
which confirms that there is a simple pole located at $N=-3i/\Lambda$ and a double pole at $N=0.$

\begin{figure}[ht]
	\centering
	\includegraphics[width=0.5\textwidth]{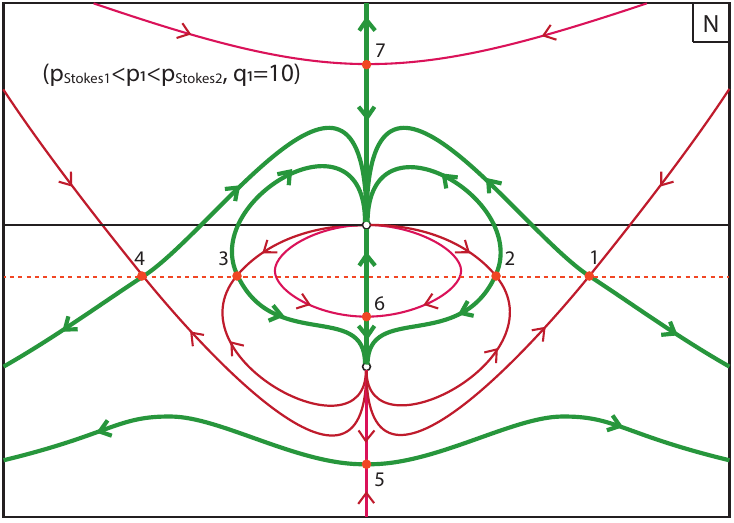}\\
 \includegraphics[width=0.45\textwidth]{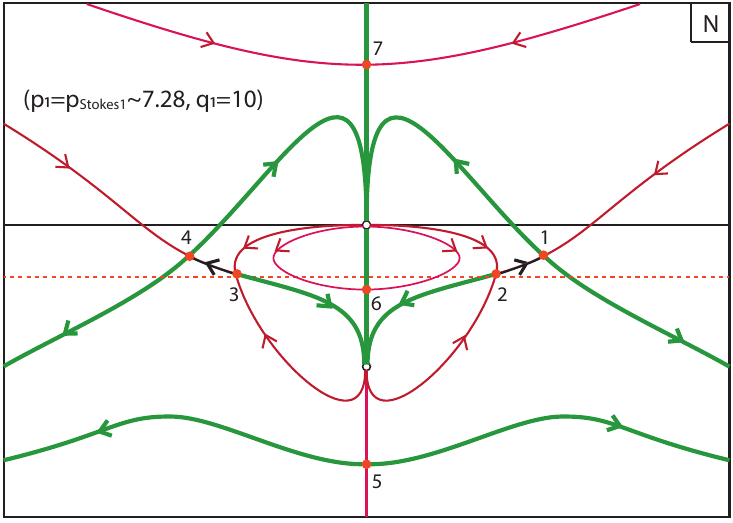}
 \includegraphics[width=0.45\textwidth]{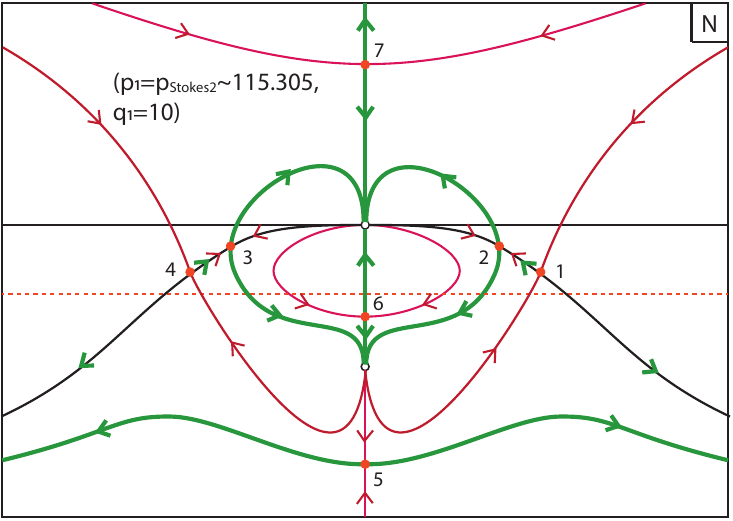}\\
 \includegraphics[width=0.45\textwidth]{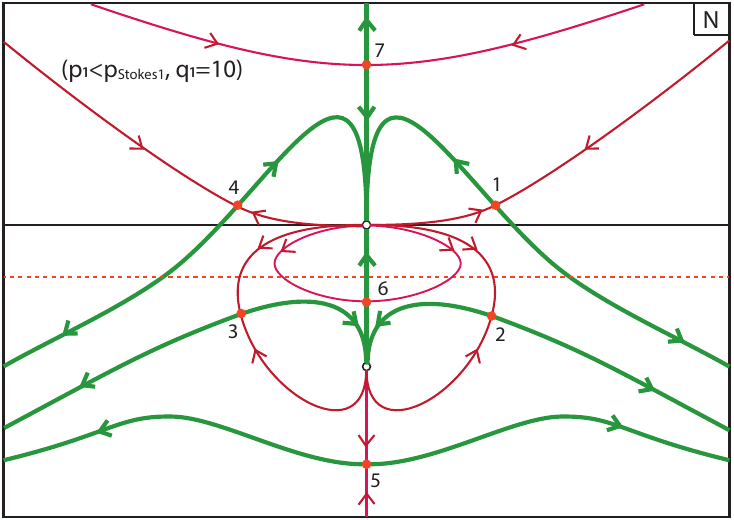}
 \includegraphics[width=0.45\textwidth]{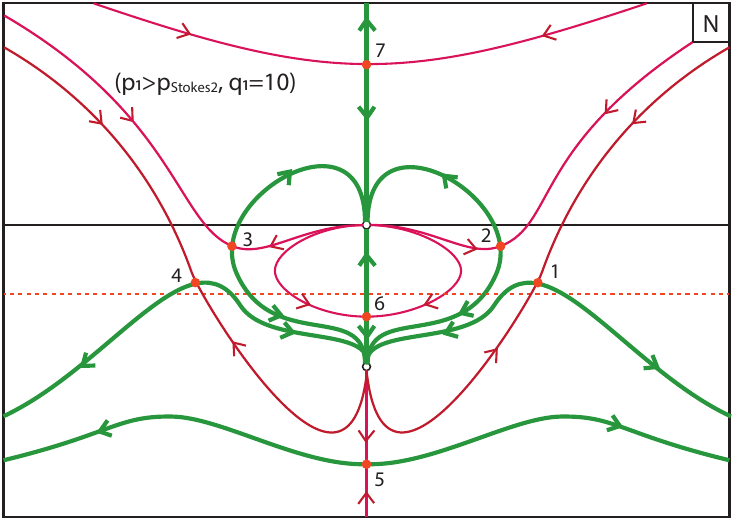}
	\caption{Flow lines for the Taub-Bolt action \eqref{Boltaction}. The top panel includes the case where the final scale factors are equal, {\it i.e.} it includes the isotropic limit. The left column then shows what happens to the flow lines as $p_1$ is decreased, to and beyond the Stokes value $p_{Stokes 1},$ while the right column shows the flow lines for increased $p_1,$ to and beyond the second Stokes value $p_{Stokes 2}.$  }
	\label{fig:posHH}
\end{figure}

\begin{figure}[ht]
	\centering
	\includegraphics[width=0.45\textwidth]{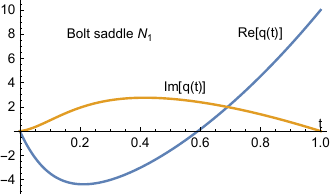}
 \includegraphics[width=0.45\textwidth]{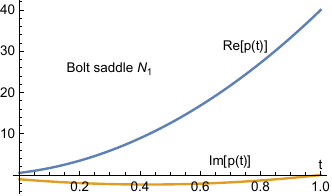}
	\caption{The typical shape of a Taub-Bolt-de Sitter instanton. The circle scale factor $q$ starts out at zero size, but the sphere radius $p$ is non-zero and complex initially. Both scale factors reach the desired real final values at $t=1.$ The parameters used here are $\Lambda=3, \, q_1=10, \, p_1=40$.}
	\label{fig:posHHsadBolt}
\end{figure}

The action admits seven stationary points in general -- for an example of the associated flow lines, see the top panel in Fig.~\ref{fig:posHH} (the conventions are the same as in Fig.~\ref{fig:stokes}). Three saddle points, labelled $5,6,7$ in the figure, are purely Euclidean and cannot describe a classical spacetime. The other four saddles represent complex metrics which come in complex conjugate pairs. They are clearly good candidate no-boundary saddles. A typical saddle point geometry is shown in Fig.~\ref{fig:posHHsadBolt}, with the same boundary conditions as those used to draw the NUT geometry shown in Fig.~\ref{fig:posHHsadNUT}. The present geometry is a complex version of the Taub-Bolt-de Sitter spacetime, with the sphere radius $p$ starting out at a non-zero complex value. Despite this, the spatial volume is initially zero, since the circle radius $q$ vanishes at $t=0.$

Which saddles contribute to the wave function depends on the integration contour chosen for the lapse integral. We can determine the appropriate choices for the integration contour by analysing the isotropic limit $q_1=p_1,$ which is included in the top sketch in Fig.~\ref{fig:posHH}. At large $N,$ we have (up to a positive numerical factor) $S_{Bolt} \sim -\frac{\Lambda}{p_1}N^3,$ which implies that the asymptotic regions of convergence are given by the wedges with angle $\theta$ comprised between the values $\pi/3 < \theta < 2\pi/3,$ $\pi < \theta<4\pi/3$ and $5\pi/3 < \theta < 2\pi.$ Since we do not want to include the Euclidean saddle $7,$ we should choose a contour that asymptotically runs between the wedges $\pi < \theta<4\pi/3$ and $5\pi/3 < \theta < 2\pi.$ What remains to be determined is where the contour passes with respect to the poles at $N=0$ and $N=-3i/\Lambda.$ If the contour passed below $N=-3i/\Lambda,$ then only the Euclidean saddle $5$ would contribute to the path integral. This leaves the two choices of passing in between the two poles, or above $N=0.$ Passing above $N=0$ would pick up only saddles $1$ and $4,$ while passing in between poles would pick up all four saddles $1,2,3,4.$ At this stage, both choices seem equally plausible.

The situation gets clarified when we look at what happens when $p_1$ is decreased, keeping $q_1$ fixed. Here one notices that the phases $\textrm{Re}(S_{Bolt})$ of saddles $1$ and $2$ (and likewise $3$ and $4$) approach each other and eventually coincide (this occurs {\it e.g.} when $p_1\approx 7.28, \, q_1=10, \, \Lambda=3$). At that stage, the two saddles become linked by a Stokes ray, see the middle left panel in Fig.~\ref{fig:posHH}. By further reducing $p_1,$ we obtain a change in the flow lines, as now the steepest ascent lines emanating from saddles $1$ and $4$ no longer link up to the pole at $N=-3i/\Lambda,$ but instead asymptote to the double pole at $N=0.$ This implies that if the integration contour is chosen to pass above $N=0,$ then saddles $1$ and $4$ continue to contribute to the path integral, while if the contour passes below $N=0$ they do not. As we will see below, the weighting of saddles $1$ and $4$ diverges in the limit that $p_1$ becomes small. Thus, if they were to contribute, they would cause the wave function to become non-normalisable. Even if the integral were cut off at some finite $p_1$ value, so that normalisability might be argued not to be a concern, we would find that these highly anisotropic saddle point geometries would be much more likely than isotropic and near-isotropic geometries. This does not seem to agree with what we know about the early universe. Hence we are led to choose the integration contour that passes between the two poles, and reaches the wedges $\pi < \theta<4\pi/3$ and $5\pi/3 < \theta < 2\pi$ asymptotically. This preferred contour of integration is indicated by a dashed orange line on the graphs. 

\begin{figure}[ht]
	\centering
	\includegraphics[width=0.45\textwidth]{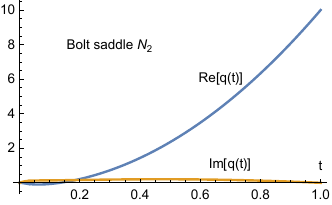}
 \includegraphics[width=0.45\textwidth]{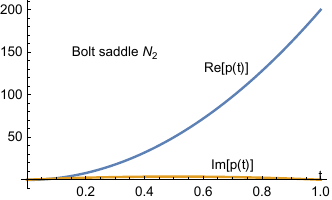}
	\caption{An example of a saddle point geometry removed from the path integral due to teh Stokes phenomenon described in the text. Here $\Lambda=3, q_1=10, p_1=200.$}
	\label{fig:posHHsadBolt2}
\end{figure}

\begin{figure}[ht]
	\centering
	\includegraphics[width=0.45\textwidth]{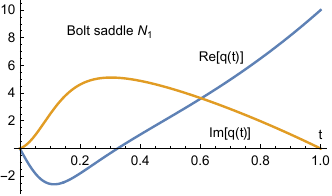}
 \includegraphics[width=0.45\textwidth]{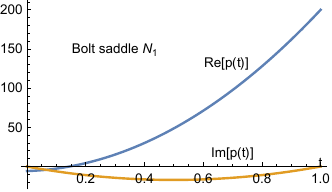}
	\caption{An example of a saddle point geometry contributing to the path integral at large squashing. Here again the final conditions are $\Lambda=3, q_1=10, p_1=200.$}
	\label{fig:posHHsadBolt1}
\end{figure}

If one now sticks to this contour of integration, it also has consequences when $p_1$ becomes large, {\it i.e.} when the final geometry is squashed in the direction of having a large sphere and a comparatively small circular direction. This is indicated by the two right-most panels in Fig.~\ref{fig:posHH}. Here one finds that a second Stokes phenomenon happens ({\it e.g.} at $p_1 \approx 115.305,\, q_1=10,\, \Lambda=3$), but this time it has the consequence of removing saddles $2$ and $3$ from the path integral. Figs.~\ref{fig:posHHsadBolt2} and \ref{fig:posHHsadBolt1} show examples of the geometries associated with saddle $2$ (which is removed from the path integral) and saddle $1$ (which contributes to the integral) at such large squashing of the geometry.

\begin{figure}[ht]
	\centering
	\includegraphics[width=0.45\textwidth]{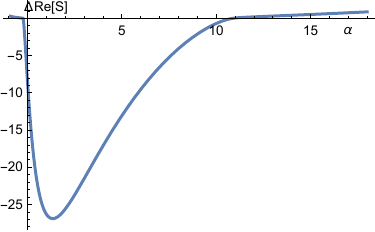}
	\caption{This graph shows the change in phase between saddles $1$ and $2.$ When the change is zero, which can be seen to occur twice, the saddles are linked by a steepest descent/ascent line, and a Stokes phenomenon occurs when the arguments of the wave function pass through the Stokes values. Here $\Lambda=3, V=125,$ and $\alpha_{Stokes 1} \approx -0.272$ while $\alpha_{Stokes 2} \approx 10.5 .$}
	\label{fig:posHHphase}
\end{figure}

Overall then, we find that at small $p_1$ only saddles $2$ and $3$ contribute to the path integral; all four saddles $1,2,3,4$ contribute in between Stokes phenomena; and solely saddles $1$ and $4$ contribute at large $p_1$. In the next section we will analyse the physical consequences of these Stokes phenomena in more detail. In concluding this section, we would simply like to verify that we have found all Stokes phenomena. To this end, we may plot the difference between the phases associated with the saddle points $1$ and $2,$ see Fig.~\ref{fig:posHHphase}. The graph uses equivalent variables to $q_1, p_1,$ which however may be physically more useful, namely the average spatial volume $V$ (rescaled by $2\pi^2$) and an anisotropy parameter $\alpha$ (previously used in \cite{Janssen:2019sex}),
\begin{align}
    V^2 = q_1 p_1^2\,, \quad \alpha = \frac{p_1}{q_1} - 1\,, \qquad p_1=V^{2/3}(1+\alpha)^{1/3}\,, \quad q_1=\left( \frac{V}{1+\alpha}\right)^{2/3}\,.
\end{align}
As the figure confirms, at constant spatial volume, there are only two instances where the phases coincide and where a Stokes phenomenon can occur, with $\alpha_{Stokes 1} \approx -0.272$ and $\alpha_{Stokes 2} \approx 10.5.$

\subsection{Dominant saddles in the wave function}

Having analysed the contributions of both NUT and Bolt types of geometries to the wave function, we may now determine the consequences for the no-boundary wave function, which we take to sum over both sets of no-boundary initial conditions, {\it i.e.} we take it to contain a sum over topologies. Let us write out the definition of the wave function more explicitly:
\begin{align}
    \Psi(q_1,p_1) = \sum_{(q_0,\Pi_{p,0})\&(p_0,\Pi_{q,0})} \int^{q_1} Dq \int^{p_1} Dp \int_{\cal C} DN \,\, e^{\frac{i}{\hbar}\left[ \int_{\cal M} \mathrm{d} ^4 x  \sqrt{-g} \left( \frac{R}{2} - \Lambda \right)  + \int_{\partial {\cal M}_{1}} \mathrm{d}^3 y \sqrt{h}K \right]}\,,
\end{align}
where we implicitly assume the minisuperspace reduction in which the curvature terms in the action depend on $q,p,N.$ Note that we also assume, as seems natural, that the two boundary conditions are included with equal weight.

Given that $\hbar$ is small, we may perform a saddle point approximation, as briefly reviewed in section \ref{sec:PL}. Of course, as discussed above, only saddles relevant to the integral can be included. And out of these, the saddle(s) with the highest weighting will exponentially dominate over saddles with a lower weighting, so that in the small $\hbar$ limit we may disregard all sub-dominant saddles.

\begin{figure}[ht]
	\centering
 \includegraphics[width=0.6\textwidth]{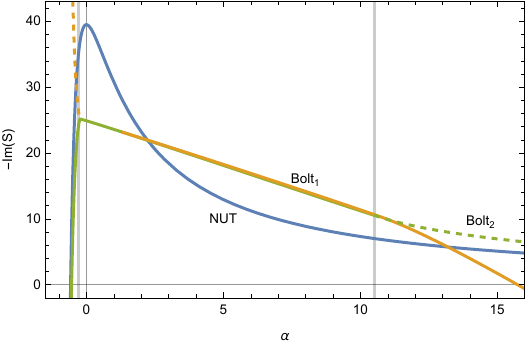}
	\caption{The weightings of various saddle points as a function of the anisotropy parameter $\alpha.$ The NUT saddle is shown in blue, and the Bolt saddles in orange (saddle $1$) and green (saddle $2$). A solid curve indicates that the saddle point in question contributes to the path integral, and a dashed curve that it does not. Here $\Lambda=3$ and the spatial volume is fixed at $V=125.$ The vertical grey lines indicate the locations of the Stokes phenomena.}
	\label{fig:posHHweightfull}
\end{figure}

In Fig.~\ref{fig:posHHweightfull} we have plotted the weighting of both the relevant NUT saddle (in blue) as well as the two types of Bolt saddles (saddle $1$ in orange and saddle $2$ in green). Note that in each case, there exist saddles with complex conjugate geometries, which have the same weighting ({\it e.g.} in the Bolt case saddle $4$ has the same weighting as saddle $1,$ and likewise for saddles $3$ and $2$). The weightings are shown at fixed spatial volume ($V=125$ and $\Lambda=3$), and as a function of the anisotropy parameter $\alpha.$ The results do not change qualitatively (and change very little quantitatively) as the final volume $V$ is increased further. Recall that $\alpha$ is defined such that $\alpha=0$ corresponds to the isotropic configuration $q_1=p_1$ for which the final hypersurface is a round $3-$sphere. As the figure shows, at $\alpha=0$ the (blue) NUT geometry dominates.

\begin{figure}[ht]
	\centering
	\includegraphics[width=0.45\textwidth]{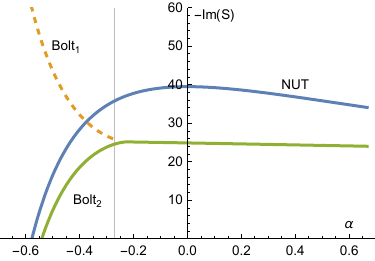}\,\,\,\,
 \includegraphics[width=0.45\textwidth]{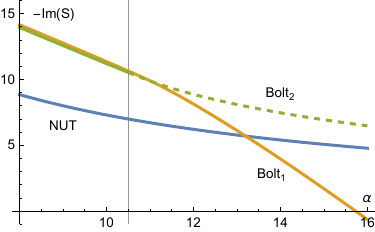}
	\caption{Zoom-ins of Fig.~\ref{fig:posHHweightfull} around the locations of the Stokes phenomena. The thin vertical lines indicate the $\alpha$ values where Stokes transitions take place.}
	\label{fig:posHHweight}
\end{figure}

At smaller values of $\alpha,$ the NUT weighting drops sharply, while it seems that the Bolt saddle $1$ comes to dominate. Its weighting rises steeply, thereby threatening to ruin the normalisability of the wave function. Moreover, if it were to contribute to the wave function, it would predict a very high probability for highly anisotropic configurations, a clear sign of an instability. This, however, does not happen, as the Stokes phenomenon (occurring at $\alpha \approx -0.272$) eliminates the Bolt saddle $1$ (and by analogy $4$) from the path integral. For this reason, we have indicated the corresponding weighting by a dashed line in the figure. A close-up of this transition is shown in the left panel of Fig.~\ref{fig:posHHweight}, from which one can see that the Stokes phenomenon occurs very close to the location where the two types of saddles have equal weight. But, as implied by the left middle panel of Fig.~\ref{fig:posHH}, saddle $2$ actually has a slightly higher weighting at the moment of the transition, since a steepest descent line links it to saddle $1$ at the Stokes value. Thus saddle $1$ is already removed while saddle $2$ still has a higher weighting, thereby preventing a potential discontinuity in the wave function (if there were no dominant NUT saddle present at those parameter values, say).

At larger values of $\alpha$ ($2.2 \lessapprox \alpha \lessapprox 13.2$) the Bolt saddle $1$ is relevant, and in fact dominant. For this range of $\alpha$ values, the Bolt topology has a higher probability than the NUT geometry. Then at even larger values of $\alpha \gtrapprox 13.2$ the NUT geometry takes over once more, and receives the highest weighting. This is because another Stokes phenomenon occurs at $\alpha\approx 10.5,$ where the Bolt saddle $2$ suddenly ceases to contribute to the wave function. If it were to contribute it would in fact remain dominant at larger $\alpha$ values, and would be more likely than the NUT geometry. However, the Stokes phenomenon removes it from the path integral. Again, the Stokes phenomenon occurs just before saddle $2$ would surpass saddle $1$ in terms of weighting, as evidenced by the steepest descent line linking saddle $1$ to saddle $2$ in the right middle panel of Fig.~\ref{fig:posHH}. Once more a potential discontinuity in the wave function is thereby avoided. And as a consequence, a topological transition from Bolt to NUT can later occur at the larger anisotropy value $\alpha \approx 13.2 .$

As we have seen, at all values of $\alpha$ (except for the phase transition at $\alpha\approx 
13.2$), the wave function is dominated by two saddle points with complex conjugate geometries (right at the phase transition there are four equally dominant saddles). This implies that, to leading order in $\hbar,$ the wave function is of the form
\begin{align}
     \Psi(q_1,p_1) \approx {\cal A} e^{\frac{i}{\hbar}\bar{S}} + {\cal A}^\star e^{-\frac{i}{\hbar}\bar{S}^\star}\,, \label{Psiapprox}
\end{align}
where $\bar{S}$ is the on-shell value of the action at one of the dominant saddle points. As one can see, being a sum of complex conjugate terms the wave function is real overall, as originally proposed by Hartle and Hawking \cite{Hartle:1983ai}. Here though the reality of the wave function stems not from the definition of the wave function as a Euclidean path integral (which we have seen not to be a tenable proposition), but rather from the symmetry of the integration contour for the lapse.

\begin{figure}[ht]
	\centering
	\includegraphics[width=0.45\textwidth]{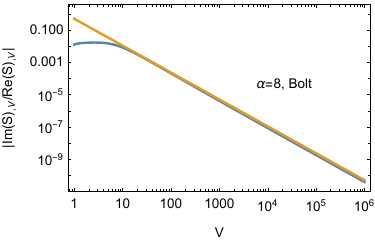}\,\,\,\,
 \includegraphics[width=0.45\textwidth]{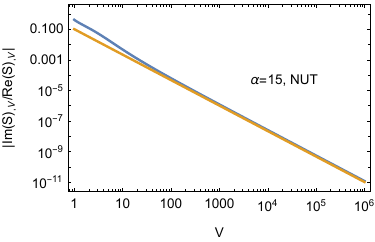}
	\caption{The scaling of the classicality condition \eqref{classy} is seen to be followed with great precision as the universe expands, both when the Bolt geometry dominates (left panel) and when the NUT geometry dominates (right panel). In both graphs the blue curves correspond to a numerical evaluation of the classicality condition, while the orange curve indicates the expected late time approximation $\propto V^{-5/3}.$}
	\label{fig:class}
\end{figure}

It turns out that the on-shell action of the dominant saddles has the same form, regardless of whether the NUT or Bolt geometry happens to dominate. In each case, we find that it is well approximated by the expression
\begin{align}
    \bar{S} = R_1(\alpha) V + \cdots \, - i \left( I_1(\alpha) + I_2(\alpha)V^{-2/3} +\cdots \right) \,,\label{sadonshell}
\end{align}
for some functions $R_i(\alpha),\, I_i(\alpha)$ that depend solely on the anisotropy. That is to say, the phase of the wave function grows in proportion to the spatial volume of the universe, while the weighting is asymptotically constant as the universe expands. A rapidly evolving phase and a slowly evolving weighting are precisely the characteristics of a WKB wave function, which may be associated with classical behaviour. Thus, we recover the prediction of classical spacetime as the universe expands out of its no-boundary initial state. We can make this statement more quantitative: Eq.~\eqref{sadonshell} implies that the ratio of the change of the weighting and the change in the phase is given by
\begin{align}
    \abs{\frac{\partial_V \textrm{Im}(\bar{S})}{\partial_V \textrm{Re}(\bar{S})}} \propto V^{-5/3}\,, \label{classy}
\end{align}
up to subdominant terms, and where we have made use of the fact that with increasing volume derivatives with respect to $\alpha$ are negligible compared to volume derivatives. Fig.~\ref{fig:class} shows that this scaling of the classicality condition is obeyed to very high precision, both in the case where the Bolt geometry is dominant (the example in the figure is taken at $\alpha=8$) and where the NUT geometry dominates ($\alpha=15$ in the figure).

A final word about the normalisability of the wave function: at small $\alpha$ the weighting of the dominant NUT saddles drops sharply, so that the wave function is evidently normalisable there (at fixed spatial volume $V$). However, at large $\alpha$ the weighting of the NUT saddles approaches zero, so that the normalisability depends on the scaling of the prefactor ${\cal A}$ in Eq.~\eqref{Psiapprox}, see also the discussion in \cite{Janssen:2019sex}. Calculating this prefactor goes beyond the aims of the present work, and is left for future investigations.

\section{Tunneling wave function} \label{sec:tunnel}

We will just add a brief remark about the tunneling wave function \cite{Vilenkin:1982de}, which is often seen as an alternative to the no-boundary wave function -- for a more detailed discussion, see {\it e.g.} \cite{Vilenkin:2018dch}. As discussed by many authors previously \cite{Bousso:1996au,Conti:2015ruo,Feldbrugge:2018gin}, the tunneling wave function suffers from an instability as it gives high weighting to anisotropic spacetimes, compared to more symmetric, isotropic configurations. This property also renders the wave function non-normalisable and thus ill defined. 

\begin{figure}[ht]
	\centering
	\includegraphics[width=0.5\textwidth]{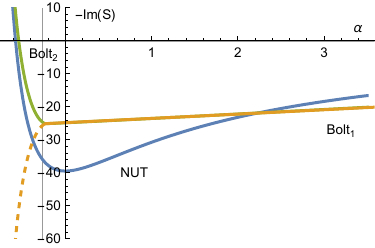}
	\caption{The weighting of various saddle point geometries in the tunneling wave function, as a function of the anisotropy parameter $\alpha$. Here $\Lambda=3, V=125.$}
	\label{fig:posTweight}
\end{figure}

We can illustrate how this instability manifests itself in the present context by implementing boundary conditions similar to those called $2$ and $3$ in section \ref{sec:lapse}, but which differ by the sign of the momentum assignments. This results in NUT and Bolt actions that are complex conjugates of those derived above, explicitly
\begin{align}
    \frac{1}{2\pi^2}S_{NUT}^T= -\frac{q_1p_1}{N} - i q_1 + (4-\frac{\Lambda}{3}p_1)N +i\frac{\Lambda}{3}N^2 \,,
\end{align}
\begin{align}
    \frac{1}{2\pi^2}S_{Bolt}^T = \left(4-\frac{\Lambda}{3} \right)N -\frac{p_1 q_1}{N}-\frac{4 \Lambda ^2 N^6 -24 i \Lambda  N^5-N^4 \left(\Lambda ^2 p_1^2+36\right)+6 \Lambda  N^2 p_1^2 q_1-9 p_1^2 q_1^2}{12 N^2 p_1 (\Lambda  N -3 i)}\,.
\end{align}
As a consequence, the weighting of the saddle point geometries is reversed, see Fig.~\ref{fig:posTweight}, as long as the same saddle points contribute to the path integral. 

\begin{figure}[ht]
	\centering
	\includegraphics[width=0.5\textwidth]{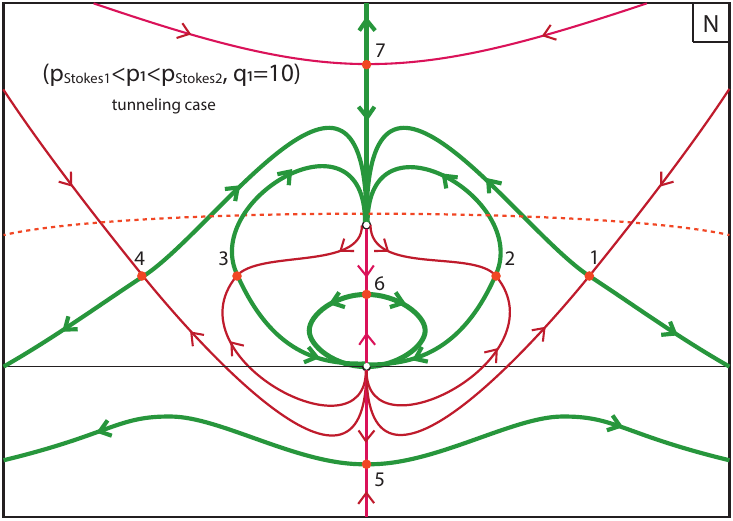}
	\caption{Flow lines of Bolt saddle points for near-isotropic final conditions $q_1 \approx p_1,$ when tunneling boundary conditions are imposed instead of no-boundary conditions.}
	\label{fig:posT}
\end{figure}

\begin{figure}[ht]
	\centering
	\includegraphics[width=0.45\textwidth]{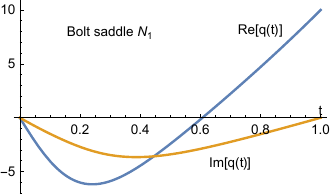}
 \includegraphics[width=0.45\textwidth]{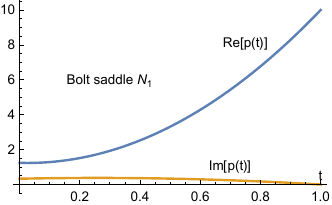}
	\caption{The dominant saddle point in the tunneling wave function for isotropic final conditions $\Lambda=3, q_1=10, p_1=10$. Note that $p$ does not start at zero, and in fact the evolutions of the two scale factors $q$ and $p$ is significantly different throughout, even though the boundary conditions are symmetric. This is in clear contrast to the no-boundary case.}
	\label{fig:posTsadBolt}
\end{figure}

Here we will focus on the Bolt geometries near $\alpha=0,$ {\it i.e.} for near-isotropic final conditions. The corresponding flow lines are shown in Fig.~\ref{fig:posT}. The expectation would be that the tunneling wave function should be defined by summing over Lorentzian geometries, so that one would expect the lapse integral to run over the real $N$ line. One then has to make the choice of whether to pass above or below the pole in the action at $N=0.$ If one passes below, one picks up the Euclidean saddle point $5,$ while if one passes just above, it turns out that the dominant saddle is saddle $6,$ which is again Euclidean. In both cases one does not obtain the prediction of a classical spacetime. Hence one is forced to take the contour above the pole at $N=3i/\Lambda,$ {\it i.e.} one is forced to take a non-Lorentzian integration contour. In this case saddles $1$ and $4$ are indeed picked up, and dominate the wave function. The corresponding geometry is shown in Fig.~\ref{fig:posTsadBolt}. What is surprising, is that this geometry is not isotropic at all, even when the final conditions are.

In this setting, Stokes phenomena once again exist at the values $\alpha=-0.272$ and $\alpha=10.5,$ but they do not change the general conclusion: as Fig.~\ref{fig:posTweight} clearly shows, anisotropic configurations are preferred, rendering the tunneling wave function unstable and physically unacceptable.

\section{Discussion} \label{sec:disc}

The biaxial Bianchi IX model offers an interesting example of a quantum cosmological model where different topologies contribute to the wave function. It thus serves as a useful toy model for studying the no-boundary wave function, and gravitational path integrals more generally. In this paper, building on earlier treatments of the same system, we have shown that Stokes phenomena occur, in which certain saddle points are made irrelevant to the path integral. Understanding these Stokes phenomena is crucial for determining both the mathematical properties, and the physical consequences, of the wave function. In particular, the Stokes phenomena ensure that the wave function remains stable and normalisable at small values of the anisotropy parameter $\alpha,$ while they imply that at large $\alpha$ the NUT geometries dominate over the Bolt geometries, even though the latter have higher weighting and would naively be expected to dominate.

We should emphasise that the phase transition at large anisotropy $\alpha$ does not simply occur because one saddle receives a higher weighting than another, but rather that at smaller anisotropy already the real parts of the action for two saddle points become identical, after which a change in the flow lines occurs with the consequence that a saddle point loses relevance. This realisation ties up a few loose ends present in earlier treatments, and justifies assumptions made there \cite{Janssen:2019sex}.

Our work has a bearing on a number of conceptual issues that one faces when dealing with gravitational path integrals: it shows that a sum over topologies can be implemented in gravitational path integrals, with phase transitions occurring between topologies. The latter statement requires a proper definition of all integration contours, especially that of the lapse function. In this respect, our results confirm earlier findings that in the no-boundary context neither a sum over purely Lorentzian, nor a sum over purely Euclidean metrics can be taken as fundamental. Since one is thus forced to work with complex metrics, the question arises as to which complex metrics should be allowed in path integrals. All? Or only those satisfying certain stability properties? These issues have started being explored in recent works \cite{Witten:2021nzp,Lehners:2021mah,Jonas:2022uqb,Lehners:2022xds}, and it would be of interest to extend these techniques to the present minisuperspace model also, especially since these works have demonstrated that a restriction to allowable metrics has observable consequences \cite{Hertog:2023vot,Lehners:2023pcn}, such as predicting a comparatively small universe.

Various directions for future exploration suggest themselves:  how might one define the lapse integral in general, rather than in a case by case manner? Could one include further topologies, not yet included here, but leading to the same final hypersurface \cite{Daughton:1998aa}? Can one use the present model as a case study for the holographic no-boundary proposal \cite{Hertog:2011ky} and, with $\Lambda$ taken negative, to better understand the AdS/CFT correspondence when the sum over bulk metrics includes different topologies? Could one understand the CFT equivalent of the Stokes phenomena in that context?

\vspace{0.2cm}
\noindent {\bf Acknowledgments}

I gratefully acknowledge the support of the European Research Council via the ERC Consolidator Grant CoG 772295 ``Qosmology''.

\bibliographystyle{JHEP2}
{\renewcommand{\baselinestretch}{1.15}
\bibliography{anisorefs}}

\providecommand{\url}[1]{#1}\providecommand{\href}[2]{#2}\begingroup\raggedright\begin{thebibliography}{10}

\bibitem{Honda:2024aro}
M.~Honda, H.~Matsui, K.~Okabayashi and T.~Terada, \emph{{Resurgence in
  Lorentzian quantum cosmology: no-boundary saddles and resummation of quantum
  gravity corrections around tunneling saddles}},
  \href{https://arxiv.org/abs/2402.09981}{{\ttfamily arXiv:2402.09981}}.

\bibitem{Ailiga:2023wzl}
M.~Ailiga, S.~Mallik and G.~Narain, \emph{{Lorentzian Robin Universe}},
  \href{https://doi.org/10.1007/JHEP01(2024)124}{\emph{JHEP} {\bfseries 01}
  (2024) 124} [\href{https://arxiv.org/abs/2308.01310}{{\ttfamily
  arXiv:2308.01310}}].

\bibitem{Dittrich:2023rcr}
B.~Dittrich and J.~Padua-Arg\"uelles, \emph{{Lorentzian quantum cosmology from
  effective spin foams}},  \href{https://arxiv.org/abs/2306.06012}{{\ttfamily
  arXiv:2306.06012}}.

\bibitem{Matsui:2023tkw}
H.~Matsui, \emph{{Summing over non-singular paths in quantum cosmology}},
  \href{https://doi.org/10.1088/1361-6382/ad1fc9}{\emph{Class. Quant. Grav.}
  {\bfseries 41} (2024) 055005}
  [\href{https://arxiv.org/abs/2304.12024}{{\ttfamily arXiv:2304.12024}}].

\bibitem{Gielen:2022yez}
S.~Gielen and E.~Nash, \emph{{Quantum cosmology of pure connection general
  relativity}}, \href{https://doi.org/10.1088/1361-6382/acccca}{\emph{Class.
  Quant. Grav.} {\bfseries 40} (2023) 115009}
  [\href{https://arxiv.org/abs/2212.06198}{{\ttfamily arXiv:2212.06198}}].

\bibitem{Matsui:2022lfj}
H.~Matsui, S.~Mukohyama and A.~Naruko, \emph{{No smooth spacetime in Lorentzian
  quantum cosmology and trans-Planckian physics}},
  \href{https://doi.org/10.1103/PhysRevD.107.043511}{\emph{Phys. Rev. D}
  {\bfseries 107} (2023) 043511}
  [\href{https://arxiv.org/abs/2211.05306}{{\ttfamily arXiv:2211.05306}}].

\bibitem{Jia:2022nda}
D.~Jia, \emph{{Truly Lorentzian quantum cosmology}},
  \href{https://doi.org/10.1103/PhysRevD.108.103540}{\emph{Phys. Rev. D}
  {\bfseries 108} (2023) 103540}
  [\href{https://arxiv.org/abs/2211.00517}{{\ttfamily arXiv:2211.00517}}].

\bibitem{Isichei:2022uzl}
R.~Isichei and J.a.~Magueijo, \emph{{Minisuperspace quantum cosmology from the
  Einstein-Cartan path integral}},
  \href{https://doi.org/10.1103/PhysRevD.107.023526}{\emph{Phys. Rev. D}
  {\bfseries 107} (2023) 023526}
  [\href{https://arxiv.org/abs/2210.05583}{{\ttfamily arXiv:2210.05583}}].

\bibitem{Lehners:2022mbd}
J.L.~Lehners, R.~Leung and K.S.~Stelle, \emph{{How to create universes with
  internal flux}},
  \href{https://doi.org/10.1103/PhysRevD.107.046006}{\emph{Phys. Rev. D}
  {\bfseries 107} (2023) 046006}
  [\href{https://arxiv.org/abs/2209.08960}{{\ttfamily arXiv:2209.08960}}].

\bibitem{Hawking:1981gb}
S.W.~Hawking, \emph{{The Boundary Conditions of the Universe}}, {\emph{Pontif.
  Acad. Sci. Scr. Varia} {\bfseries 48} (1982) 563}.

\bibitem{Hartle:1983ai}
J.B.~Hartle and S.W.~Hawking, \emph{{Wave Function of the Universe}},
  \href{https://doi.org/10.1103/PhysRevD.28.2960}{\emph{Phys. Rev. D}
  {\bfseries 28} (1983) 2960}.

\bibitem{Lehners:2023yrj}
J.L.~Lehners, \emph{{Review of the no-boundary wave function}},
  \href{https://doi.org/10.1016/j.physrep.2023.06.002}{\emph{Phys. Rept.}
  {\bfseries 1022} (2023) 1}
  [\href{https://arxiv.org/abs/2303.08802}{{\ttfamily arXiv:2303.08802}}].

\bibitem{Halliwell:1988ik}
J.J.~Halliwell and J.~Louko, \emph{{Steepest Descent Contours in the Path
  Integral Approach to Quantum Cosmology. 1. The De Sitter Minisuperspace
  Model}}, \href{https://doi.org/10.1103/PhysRevD.39.2206}{\emph{Phys. Rev. D}
  {\bfseries 39} (1989) 2206}.

\bibitem{Halliwell:1990tu}
J.J.~Halliwell and J.~Louko, \emph{{Steepest Descent Contours in the Path
  Integral Approach to Quantum Cosmology. 3. A General Method With Applications
  to Anisotropic Minisuperspace Models}},
  \href{https://doi.org/10.1103/PhysRevD.42.3997}{\emph{Phys. Rev. D}
  {\bfseries 42} (1990) 3997}.

\bibitem{Daughton:1998aa}
A.~Daughton, J.~Louko and R.D.~Sorkin, \emph{{Instantons and unitarity in
  quantum cosmology with fixed four volume}},
  \href{https://doi.org/10.1103/PhysRevD.58.084008}{\emph{Phys. Rev. D}
  {\bfseries 58} (1998) 084008}
  [\href{https://arxiv.org/abs/gr-qc/9805101}{{\ttfamily gr-qc/9805101}}].

\bibitem{Hartle:2008ng}
J.B.~Hartle, S.W.~Hawking and T.~Hertog, \emph{{The Classical Universes of the
  No-Boundary Quantum State}},
  \href{https://doi.org/10.1103/PhysRevD.77.123537}{\emph{Phys. Rev. D}
  {\bfseries 77} (2008) 123537}
  [\href{https://arxiv.org/abs/0803.1663}{{\ttfamily arXiv:0803.1663}}].

\bibitem{Feldbrugge:2017kzv}
J.~Feldbrugge, J.L.~Lehners and N.~Turok, \emph{{Lorentzian Quantum
  Cosmology}}, \href{https://doi.org/10.1103/PhysRevD.95.103508}{\emph{Phys.
  Rev. D} {\bfseries 95} (2017) 103508}
  [\href{https://arxiv.org/abs/1703.02076}{{\ttfamily arXiv:1703.02076}}].

\bibitem{DiazDorronsoro:2018wro}
J.~Diaz~Dorronsoro, J.J.~Halliwell, J.B.~Hartle, T.~Hertog, O.~Janssen and
  Y.~Vreys, \emph{{Damped perturbations in the no-boundary state}},
  \href{https://doi.org/10.1103/PhysRevLett.121.081302}{\emph{Phys. Rev. Lett.}
  {\bfseries 121} (2018) 081302}
  [\href{https://arxiv.org/abs/1804.01102}{{\ttfamily arXiv:1804.01102}}].

\bibitem{Feldbrugge:2018gin}
J.~Feldbrugge, J.L.~Lehners and N.~Turok, \emph{{Inconsistencies of the New
  No-Boundary Proposal}},
  \href{https://doi.org/10.3390/universe4100100}{\emph{Universe} {\bfseries 4}
  (2018) 100} [\href{https://arxiv.org/abs/1805.01609}{{\ttfamily
  arXiv:1805.01609}}].

\bibitem{Janssen:2019sex}
O.~Janssen, J.J.~Halliwell and T.~Hertog, \emph{{No-boundary proposal in
  biaxial Bianchi IX minisuperspace}},
  \href{https://doi.org/10.1103/PhysRevD.99.123531}{\emph{Phys. Rev. D}
  {\bfseries 99} (2019) 123531}
  [\href{https://arxiv.org/abs/1904.11602}{{\ttfamily arXiv:1904.11602}}].

\bibitem{Hebecker:2018ofv}
A.~Hebecker, T.~Mikhail and P.~Soler, \emph{{Euclidean wormholes, baby
  universes, and their impact on particle physics and cosmology}},
  \href{https://doi.org/10.3389/fspas.2018.00035}{\emph{Front. Astron. Space
  Sci.} {\bfseries 5} (2018) 35}
  [\href{https://arxiv.org/abs/1807.00824}{{\ttfamily arXiv:1807.00824}}].

\bibitem{Jonas:2023ipa}
C.~Jonas, G.~Lavrelashvili and J.L.~Lehners, \emph{{Zoo of axionic wormholes}},
  \href{https://doi.org/10.1103/PhysRevD.108.066012}{\emph{Phys. Rev. D}
  {\bfseries 108} (2023) 066012}
  [\href{https://arxiv.org/abs/2306.11129}{{\ttfamily arXiv:2306.11129}}].

\bibitem{Jonas:2023qle}
C.~Jonas, G.~Lavrelashvili and J.L.~Lehners, \emph{{Stability of Axion-Dilaton
  Wormholes}},  \href{https://arxiv.org/abs/2312.08971}{{\ttfamily
  arXiv:2312.08971}}.

\bibitem{Witten:2010cx}
E.~Witten, \emph{{Analytic Continuation Of Chern-Simons Theory}}, {\emph{AMS/IP
  Stud. Adv. Math.} {\bfseries 50} (2011) 347}
  [\href{https://arxiv.org/abs/1001.2933}{{\ttfamily arXiv:1001.2933}}].

\bibitem{Tanizaki:2014xba}
Y.~Tanizaki and T.~Koike, \emph{{Real-time Feynman path integral with
  Picard\textendash{}Lefschetz theory and its applications to quantum
  tunneling}}, \href{https://doi.org/10.1016/j.aop.2014.09.003}{\emph{Annals
  Phys.} {\bfseries 351} (2014) 250}
  [\href{https://arxiv.org/abs/1406.2386}{{\ttfamily arXiv:1406.2386}}].

\bibitem{DiTucci:2019dji}
A.~Di~Tucci and J.L.~Lehners, \emph{{No-Boundary Proposal as a Path Integral
  with Robin Boundary Conditions}},
  \href{https://doi.org/10.1103/PhysRevLett.122.201302}{\emph{Phys. Rev. Lett.}
  {\bfseries 122} (2019) 201302}
  [\href{https://arxiv.org/abs/1903.06757}{{\ttfamily arXiv:1903.06757}}].

\bibitem{DiTucci:2019bui}
A.~Di~Tucci, J.L.~Lehners and L.~Sberna, \emph{{No-boundary prescriptions in
  Lorentzian quantum cosmology}},
  \href{https://doi.org/10.1103/PhysRevD.100.123543}{\emph{Phys. Rev. D}
  {\bfseries 100} (2019) 123543}
  [\href{https://arxiv.org/abs/1911.06701}{{\ttfamily arXiv:1911.06701}}].

\bibitem{Jonas:2020pos}
C.~Jonas and J.L.~Lehners, \emph{{No-boundary solutions are robust to quantum
  gravity corrections}},
  \href{https://doi.org/10.1103/PhysRevD.102.123539}{\emph{Phys. Rev. D}
  {\bfseries 102} (2020) 123539}
  [\href{https://arxiv.org/abs/2008.04134}{{\ttfamily arXiv:2008.04134}}].

\bibitem{York:1972sj}
J.W.~York,~Jr., \emph{{Role of conformal three geometry in the dynamics of
  gravitation}}, \href{https://doi.org/10.1103/PhysRevLett.28.1082}{\emph{Phys.
  Rev. Lett.} {\bfseries 28} (1972) 1082}.

\bibitem{Gibbons:1976ue}
G.W.~Gibbons and S.W.~Hawking, \emph{{Action Integrals and Partition Functions
  in Quantum Gravity}},
  \href{https://doi.org/10.1103/PhysRevD.15.2752}{\emph{Phys. Rev. D}
  {\bfseries 15} (1977) 2752}.

\bibitem{Louko:1988bk}
J.~Louko, \emph{{Canonizing the Hartle-hawking Proposal}},
  \href{https://doi.org/10.1016/0370-2693(88)90008-1}{\emph{Phys. Lett. B}
  {\bfseries 202} (1988) 201}.

\bibitem{Ryan:1975jw}
M.P.~Ryan and L.C.~Shepley, \emph{{Homogeneous Relativistic Cosmologies}},
  Princeton Series in Physics, Princeton University Press, Princeton (1975).

\bibitem{Feldbrugge:2017fcc}
J.~Feldbrugge, J.L.~Lehners and N.~Turok, \emph{{No smooth beginning for
  spacetime}},
  \href{https://doi.org/10.1103/PhysRevLett.119.171301}{\emph{Phys. Rev. Lett.}
  {\bfseries 119} (2017) 171301}
  [\href{https://arxiv.org/abs/1705.00192}{{\ttfamily arXiv:1705.00192}}].

\bibitem{Feldbrugge:2017mbc}
J.~Feldbrugge, J.L.~Lehners and N.~Turok, \emph{{No rescue for the no boundary
  proposal: Pointers to the future of quantum cosmology}},
  \href{https://doi.org/10.1103/PhysRevD.97.023509}{\emph{Phys. Rev. D}
  {\bfseries 97} (2018) 023509}
  [\href{https://arxiv.org/abs/1708.05104}{{\ttfamily arXiv:1708.05104}}].

\bibitem{Anabalon:2018rzq}
A.~Anabal\'on and J.~Oliva, \emph{{Four-dimensional Traversable Wormholes and
  Bouncing Cosmologies in Vacuum}},
  \href{https://doi.org/10.1007/JHEP04(2019)106}{\emph{JHEP} {\bfseries 04}
  (2019) 106} [\href{https://arxiv.org/abs/1811.03497}{{\ttfamily
  arXiv:1811.03497}}].

\bibitem{Vilenkin:1982de}
A.~Vilenkin, \emph{{Creation of Universes from Nothing}},
  \href{https://doi.org/10.1016/0370-2693(82)90866-8}{\emph{Phys. Lett. B}
  {\bfseries 117} (1982) 25}.

\bibitem{Vilenkin:2018dch}
A.~Vilenkin and M.~Yamada, \emph{{Tunneling wave function of the universe}},
  \href{https://doi.org/10.1103/PhysRevD.98.066003}{\emph{Phys. Rev. D}
  {\bfseries 98} (2018) 066003}
  [\href{https://arxiv.org/abs/1808.02032}{{\ttfamily arXiv:1808.02032}}].

\bibitem{Bousso:1996au}
R.~Bousso and S.W.~Hawking, \emph{{Pair creation of black holes during
  inflation}}, \href{https://doi.org/10.1103/PhysRevD.54.6312}{\emph{Phys. Rev.
  D} {\bfseries 54} (1996) 6312}
  [\href{https://arxiv.org/abs/gr-qc/9606052}{{\ttfamily gr-qc/9606052}}].

\bibitem{Conti:2015ruo}
G.~Conti, T.~Hertog and E.~van~der~Woerd, \emph{{Holographic Tunneling Wave
  Function}}, \href{https://doi.org/10.1007/JHEP12(2015)025}{\emph{JHEP}
  {\bfseries 12} (2015) 025}
  [\href{https://arxiv.org/abs/1506.07374}{{\ttfamily arXiv:1506.07374}}].

\bibitem{Witten:2021nzp}
E.~Witten, \emph{{A Note On Complex Spacetime Metrics}},
  \href{https://arxiv.org/abs/2111.06514}{{\ttfamily arXiv:2111.06514}}.

\bibitem{Lehners:2021mah}
J.L.~Lehners, \emph{{Allowable complex metrics in minisuperspace quantum
  cosmology}}, \href{https://doi.org/10.1103/PhysRevD.105.026022}{\emph{Phys.
  Rev. D} {\bfseries 105} (2022) 026022}
  [\href{https://arxiv.org/abs/2111.07816}{{\ttfamily arXiv:2111.07816}}].

\bibitem{Jonas:2022uqb}
C.~Jonas, J.L.~Lehners and J.~Quintin, \emph{{Uses of complex metrics in
  cosmology}}, \href{https://doi.org/10.1007/JHEP08(2022)284}{\emph{JHEP}
  {\bfseries 08} (2022) 284}
  [\href{https://arxiv.org/abs/2205.15332}{{\ttfamily arXiv:2205.15332}}].

\bibitem{Lehners:2022xds}
J.L.~Lehners, \emph{{Allowable complex scalars from Kaluza-Klein
  compactifications and metric rescalings}},
  \href{https://doi.org/10.1103/PhysRevD.107.046004}{\emph{Phys. Rev. D}
  {\bfseries 107} (2023) 046004}
  [\href{https://arxiv.org/abs/2209.14669}{{\ttfamily arXiv:2209.14669}}].

\bibitem{Hertog:2023vot}
T.~Hertog, O.~Janssen and J.~Karlsson, \emph{{Kontsevich-Segal Criterion in the
  No-Boundary State Constrains Inflation}},
  \href{https://doi.org/10.1103/PhysRevLett.131.191501}{\emph{Phys. Rev. Lett.}
  {\bfseries 131} (2023) 191501}
  [\href{https://arxiv.org/abs/2305.15440}{{\ttfamily arXiv:2305.15440}}].

\bibitem{Lehners:2023pcn}
J.L.~Lehners and J.~Quintin, \emph{{A small Universe}},
  \href{https://doi.org/10.1016/j.physletb.2024.138488}{\emph{Phys. Lett. B}
  {\bfseries 850} (2024) 138488}
  [\href{https://arxiv.org/abs/2309.03272}{{\ttfamily arXiv:2309.03272}}].

\bibitem{Hertog:2011ky}
T.~Hertog and J.~Hartle, \emph{{Holographic No-Boundary Measure}},
  \href{https://doi.org/10.1007/JHEP05(2012)095}{\emph{JHEP} {\bfseries 05}
  (2012) 095} [\href{https://arxiv.org/abs/1111.6090}{{\ttfamily
  arXiv:1111.6090}}].

\end{thebibliography}\endgroup

\end{document}